\documentclass[12pt]{article}
\usepackage{amssymb}
\usepackage{graphicx}
\usepackage[cp1251]{inputenc}
\usepackage{rotating}

\begin{document}
 \noindent {\footnotesize\it Astrophysical Bulletin, 2025, Vol. 80, No. 2}
 \newcommand{\dif}{\textrm{d}}

 \noindent
 \begin{tabular}{llllllllllllllllllllllllllllllllllllllllllllll}
 & & & & & & & & & & & & & & & & & & & & & & & & & & & & & & & & & & & & & &\\\hline\hline
 \end{tabular}

  \vskip 0.5cm
  \bigskip
 \bigskip
\centerline{\large\bf  Age Estimation of the Radcliffe Wave from Open Star Clusters}

 \bigskip
 \bigskip
  \centerline { 
   V. V. Bobylev\footnote [1]{bob-v-vzz@rambler.ru}, N. R. Ikhsanov, and A. T. Bajkova}
 \bigskip
 \centerline{\small\it Pulkovo Astronomical Observatory, Russian Academy of Sciences, St. Petersburg, 196140 Russia}
 \bigskip
 \bigskip
{Abstract---Four samples of open star clusters (OSCs) with average ages of 5.2, 18.6, 40, and 61 Myr have been analyzed. The selection of these OSCs was carried out from a narrow region inclined to the galactic axis y at an angle of 25$^\circ$. The spectral analysis of the vertical positions and velocities of the selected clusters showed that the Radcliffe wave is associated with OSCs no older than 30 Myr. The following estimates of the Radcliffe wave characteristics were obtained for the OSCs with an average age of 5.2 Myr: $z_{max}=117\pm12$ pc with the wavelength $\lambda=4.55\pm0.14$ kpc, the vertical velocity disturbance amplitude
$W_{max}=4.86\pm0.19$ km s$^{-1}$ with the wavelength $\lambda=1.74\pm0.08$ kpc. For the OSCs with an average
age of 18.6 Myr, the estimates are as follows: $z_{max} = 54\pm5$ pc and $\lambda=6.30\pm0.12$ kpc, the vertical
velocity disturbance amplitude $W_{max}=7.90\pm0.16$ km s$^{-1}$ and $\lambda=0.83\pm0.11$ kpc. The radial motion
of the Radcliffe wave away from the galactic center has been confirmed. The velocity of such movement is
10 pc Myr$^{-1}$. In our opinion, the spatial distribution of OSCs younger than 30 Myr does not contradict
the hypothesis of the association of the Radcliffe wave with the impact of shock waves from supernova
explosions that arose on an extended front comparable in scale to the entire wave, that is, about 2 kpc in
size.
 }

\bigskip
\section{Introduction}
The Radcliffe wave was first discovered by Alves et al.  (2020) while studying the spatial distribution of a large sample of molecular clouds located in the solar neighborhood. The authors identified a narrow chain of clouds stretched almost in a line with a length of
about 2.7 kpc in the galactic plane $xy$. The main feature of this structure, called the Radcliffe wave, is the wave-like nature of the distribution in the vertical direction. According to Alves et al. (2020), the wave has a fading character, and the maximum value of the
coordinate $z\sim160$~pc is observed in the immediate vicinity of the Sun.

The wave-like behavior of vertical coordinates is manifested in the distribution of the interstellar dust (Lallement et al., 2022; Edenhofer et al., 2024), molecular clouds (Zucker et al., 2023), masers, and radio stars (Bobylev et al. 2022), T Tauri stars (Li and Chen, 2022), massive OB stars (Donada and Figueras, 2021; Thulasidharan et al., 2022), and young open star clusters (Donada and Figueras, 2021).

Recently, with the use of OSCs, interesting results have been obtained about the nature of the vertical
velocities of objects in the Radcliffe wave (Alonso-Santiago et al., 2024; Bobylev and Bajkova, 2024;
Konietzka et al., 2024; Zhu et al., 2024). In this case, different catalogs with the data on OSCs were
used: the papers of Hunt and Reffert (2023) as well as Cantat-Gaudin et al. (2020). As a result, it was
established that the vertical velocities of young OSCs in the Radcliffe wave have a wave-like character with
a maximum velocity of 5--15~km s$^{-1}$ . In addition to vertical oscillations, Konietzka et al. (2024) found
evidence that the Radcliffe wave is drifting radially outward fromthe galactic center at a velocity of about
5~km s$^{-1}$ . Zhu et al. (2024) modeled the changes in wave parameters over the past 40~Myr.

To date, a number of hypotheses have been proposed to explain the origin of the Radcliffe wave. Fleck (2020) associates the origin of the Radcliffe wave to the Kelvin--Helmholtz instability which arises in the Galactic disk due to the difference in rotation velocities of the dark matter halo and the disk. The hypothesis of the influence of an external impactor on the galactic disk is discussed (Thulasidharan et al., 2022). The impactor could be a dwarf galaxy or a globular cluster. The passage of such a massive body through the galactic disk can cause wave disturbances that propagate across the disk.

Li et al. (2024) modeled the three-dimensional evolution of the local interstellar gas and showed that the gravitational influence of the Galaxy stretches the Radcliffe wave to almost twice its current length on a time scale of 45 Myr. The modeling also shows the
formation of new filaments and merging of existing filaments.

Marchal and Martin (2023) associate the origin of the Radcliffe wave to the evolution of the North
Polar Spur. They believe that the narrow chain of the matter (from which the Radcliffe wave formed) arose
at the outer edge of the North Polar Spur as a result of the impact of shock waves from several supernovae and their stellar winds. A closely related hypothesis is that developed by Konietzka et al. (2024) about the connection between the Radcliffe wave and the
evolution of the Local Bubble. In any case, the shock waves here were generated in the past by supernova
explosions in the Scorpius--Centaurus (Sco--Cen) OB association closest to the Sun. However, there is currently no generally accepted hypothesis for the formation of the Radcliffe wave.

The aim of this work is to estimate the parameters of the Radcliffe wave based on the coordinates and velocities of the OSCs of different ages and to establish the upper age limit of the OSCs that still belong to this wave. The Hunt and Reffert (2023) catalog is used, in which the kinematic characteristics of the clusters are taken from the Gaia\,DR3 (Vallenari et al., 2023), and there are also estimates of the individual age of the OSCs.

   \section{METHOD}
In this paper, we use a rectangular coordinate system $x, y, z,$ in which the $x$ axis is directed from the
Sun to the center of the Galaxy, the direction of the $y$ axis coincides with the direction of rotation of the
Galaxy, and the $z$ axis is directed to the north galactic pole. We also use the galactocentric rectangular
coordinate system $X,Y,Z,$ in which the $X$ axis is directed from the center of the Galaxy to the Sun, the direction of the $Y$ axis coincides with the direction of rotation of the Galaxy, and the $Z$ axis is directed to the north galactic pole. Thus, in these two coordinate systems only the directions of the x and $X$ axes differ.

The orientation of the Radcliffe wave with respect to the y and $Y$ axes differs only in sign. For example,
in the $xy$ coordinate system, the transition to the hatched $y'$ axis when rotating by the angle $\beta$ is carried
out according to the formula
$$y'=y\cos\beta+x\sin\beta.$$

To study the periodic structure in the coordinates and velocities of stars, Bobylev et al. (2022) proposed
using the spectral analysis based on the standard
Fourier transform of the original sequence $z(y')$:
\begin{equation}
 \renewcommand{\arraystretch}{2.0}
 \begin{array}{lll}
 \displaystyle
 F(z(y'))=\int z(y')e^{-j 2\pi y'/\lambda}=U(\lambda)+jV(\lambda)=A(\lambda)e^{j\varphi(\lambda)},
 \label{F}
 \end{array}
\end{equation}
where
 $A(\lambda)=\sqrt{U^2(\lambda)+V^2(\lambda)}$ is the amplitude of the spectrum, and $\varphi(\lambda)=\arctan(V(\lambda)/U(\lambda))$ is the spectrum phase. The characteristic feature of
the present approach is the search for not just a monochromatic wave with a constant amplitude, but
a wave that most accurately describes the initial data, the spectrum of which coincides with the main peak
(lobe) of the calculated spectrum in the wavelength range from $\lambda_{min}$ to $\lambda_{max}$  (within these boundaries, the spectrum smoothly decreases starting from the maximum value, and outside it begins to increase).

As a result, we have the desired smooth curve approximating the initial data, which is calculated
using the formula for the inverse Fourier transform in the wavelength range we have defined:
\begin{equation}
 z(y')= 2k\int^{\lambda_{max}}_{\lambda_{min}} A(\lambda)\cos\biggl( {2\pi y'\over\lambda} +
  \varphi(\lambda) \biggr) d\lambda,
  \label{Z}
\end{equation}
where $k$  is the coefficient calculated from the condition of the minimum residual.

   \section{DATA}
In this paper, we use the Hunt and Reffert (2023) catalog containing 4780 OSCs. For all the OSCs
in this catalog, there are estimates of age, heliocentric distance, and kinematic characteristics. Average
radial velocities calculated from the Gaia\,DR3 (Vallenari et al., 2023) are available for a large number of
OSCs, although, not for all of them.

In this paper, the selection of clusters is carried out
in a region bounded by two lines in the $xy$ plane:
\begin{equation}
 \begin{array}{lll}
 x= y\tan\beta-0.65, \\
 x= y\tan\beta+0.00,
 \end{array}
 \label{xy-0.65}
\end{equation}
where the angle $\beta=25^\circ$  was chosen so that the wave manifested itself in the best possible way. The center
of the selection region does not pass through the origin of coordinates but intersects the $x$ axis at the
point $x_1$, therefore, the transition to the $y'$ axis was performed according to the formula:
\begin{equation}
 y'= y\cos\beta+(x-x_1)\sin\beta,
 \label{y'-dx}
\end{equation}
where $x_1=-0.28$~kpc.

When forming the samples, we distributed the OSCs by age:

1) younger than 10 Myr;

2) in the interval of 10--30 Myr;

3) in the interval of 30--50 Myr;

4) from 50 to 70 Myr.

After selection according to the boundary conditions from formula (3), the first sample included 219 OSCs
with an average age of 5.2 Myr, of which the radial velocities are given for 211 OSCs in the Hunt and
Reffert (2023) catalog. The second sample contains 138 OSCs (with an average age of 16.6 Myr), of
which the radial velocities are available for 129. The third sample includes 93 OSCs with an average age
of 40 Myr, of which the radial velocities are available for 92. In the fourth sample of 112 OSCs (with
an average age of 61 Myr), the radial velocities are available for 109. Figure~1 shows the distribution of
the samples of OSCs selected for analysis on the $XY$ galactic plane.

\begin{figure}[t]
{ \begin{center}
  \includegraphics[width=0.99\textwidth]{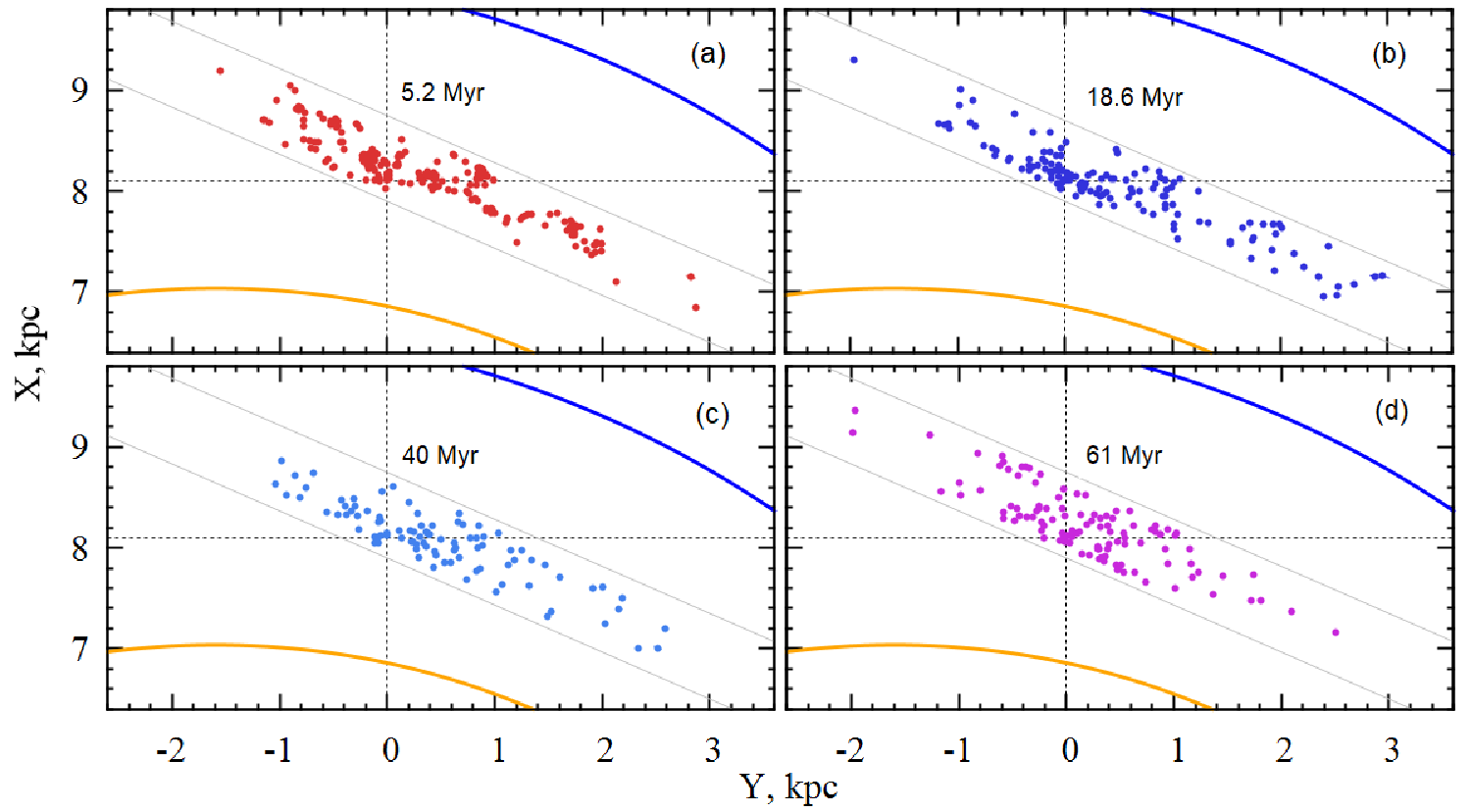}
  \caption{
  Distribution of OSCs projected onto the Galactic $XY$ plane for the samples with average ages of 5.2~Myr (a), 18.6~Myr (b),
40~Myr (c), and 61~Myr (d). The blue and orange lines show the sections of the spiral arms closest to the Sun, and the thin gray lines indicate the boundaries of the OSC selection region.
 }
 \label{f1-XY}
\end{center}}
\end{figure}

   \section{RESULTS}
Analysis of the coordinates of the OSC sample with an average age of 5.2 Myr yielded an estimate
of the maximum value of the amplitude of vertical disturbances $z_{max}$ with the wavelength of these disturbances $\lambda$:
  \begin{equation}
 \label{sol-z5}
 \begin{array}{lll}
   z_{max}= 117\pm12~\hbox{pc},\\
   \lambda= 4.55\pm0.14~\hbox{kpc}.
 \end{array}
 \end{equation}
While based on the vertical velocities of this OSC sample, we obtained an estimate of the maximum of
their disturbance velocity $W_{max}$  with the wavelength of these disturbances $\lambda$:
  \begin{equation}
 \label{sol-W5}
 \begin{array}{lll}
   W_{max}= 4.86\pm0.19~\hbox{km s$^{-1}$},\\
   \lambda= 1.74\pm0.08~\hbox{kpc},
 \end{array}
 \end{equation}
The results of the spectral analysis of the positions and velocities of these OSCs are shown in Fig. 2.

\begin{figure}[t]
{ \begin{center}
  \includegraphics[width=0.85\textwidth]{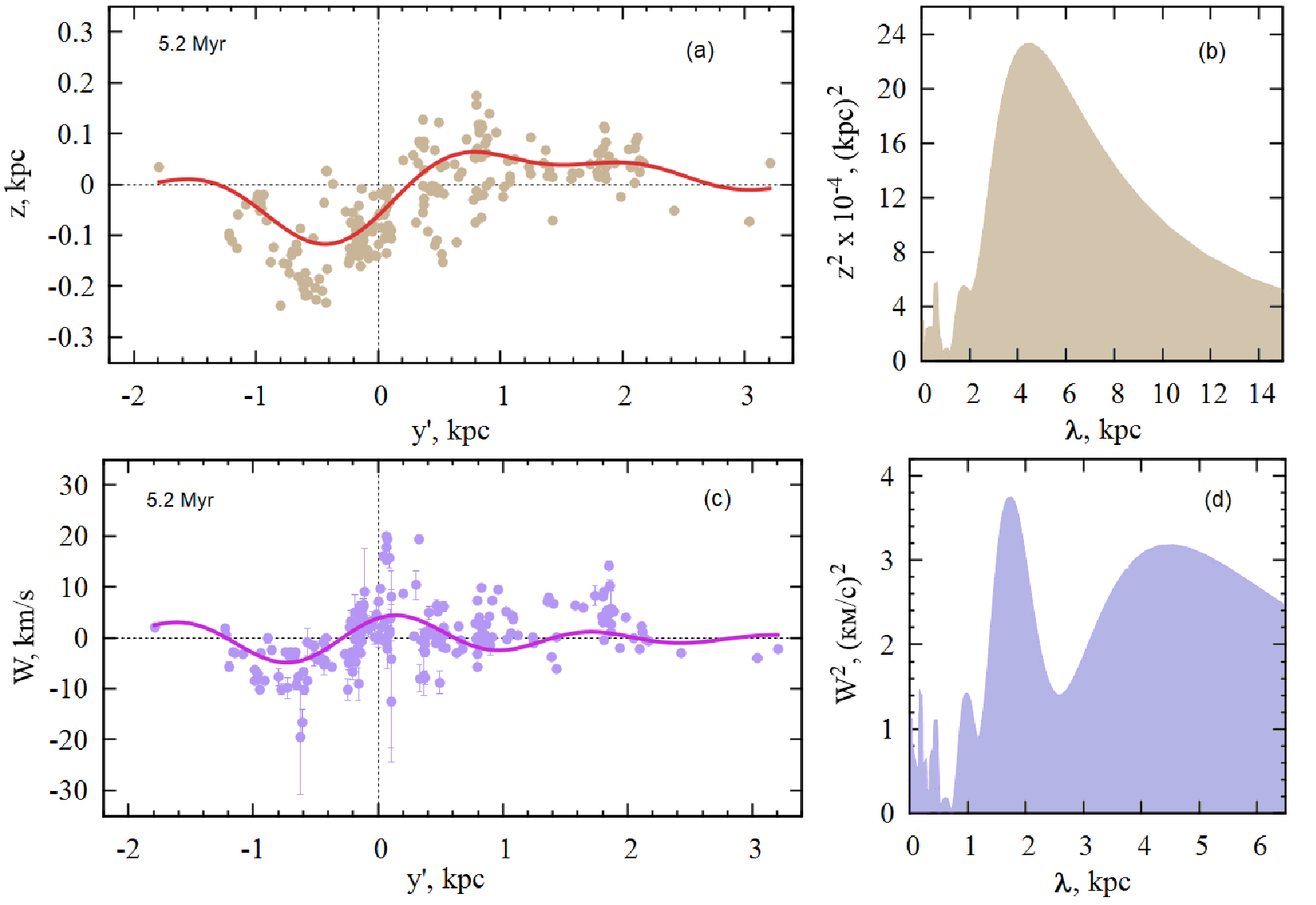}
  \caption{
 $z$ coordinates of the OSCs from the sample with an average age of 5.2 Myr depending on the distance $y'$~(a) and their power
spectrum(b), the vertical velocities $W$ of these OSCs as a function of the distance $y'$~(c), and their power spectrum(d). The periodic curves shown as solid bold lines reflect the results of the spectral analysis; for details, see the text.
 }
 \label{f2-5Myr}
\end{center}}
\end{figure}
\begin{figure}[t]
{ \begin{center}
  \includegraphics[width=0.85\textwidth]{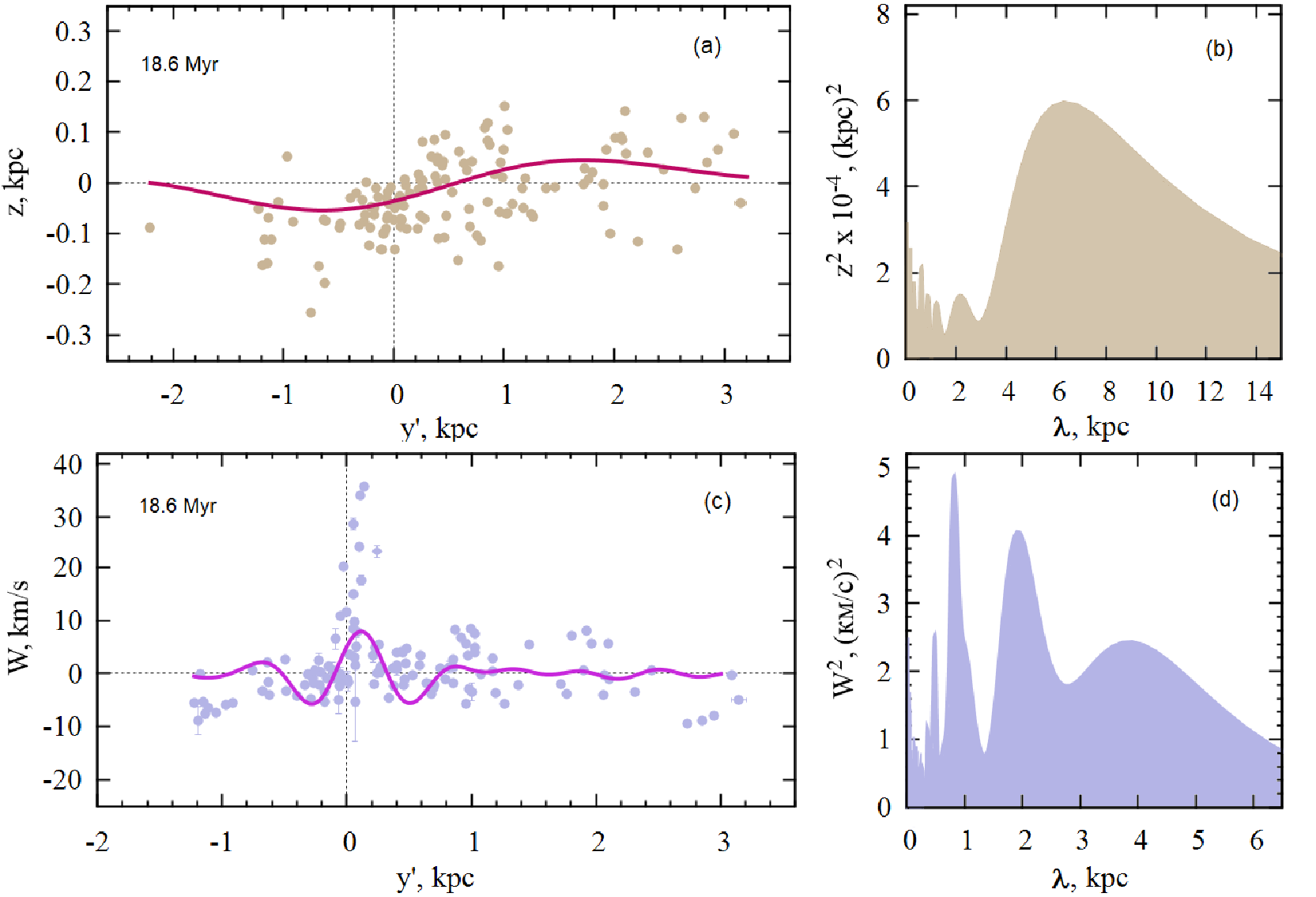}
  \caption{
 $z$ coordinates of the OSCs from the sample with an average age of 18.6 Myr depending on the distance $y'$~(a) and their power
spectrum(b), the vertical velocities $W$ of these OSCs as a function of the distance $y'$~(c), and their power spectrum(d). The periodic curves shown as solid bold lines reflect the results of the spectral analysis.
 }
 \label{f3-18Myr}
\end{center}}
\end{figure}
\begin{figure}[t]
{ \begin{center}
  \includegraphics[width=0.95\textwidth]{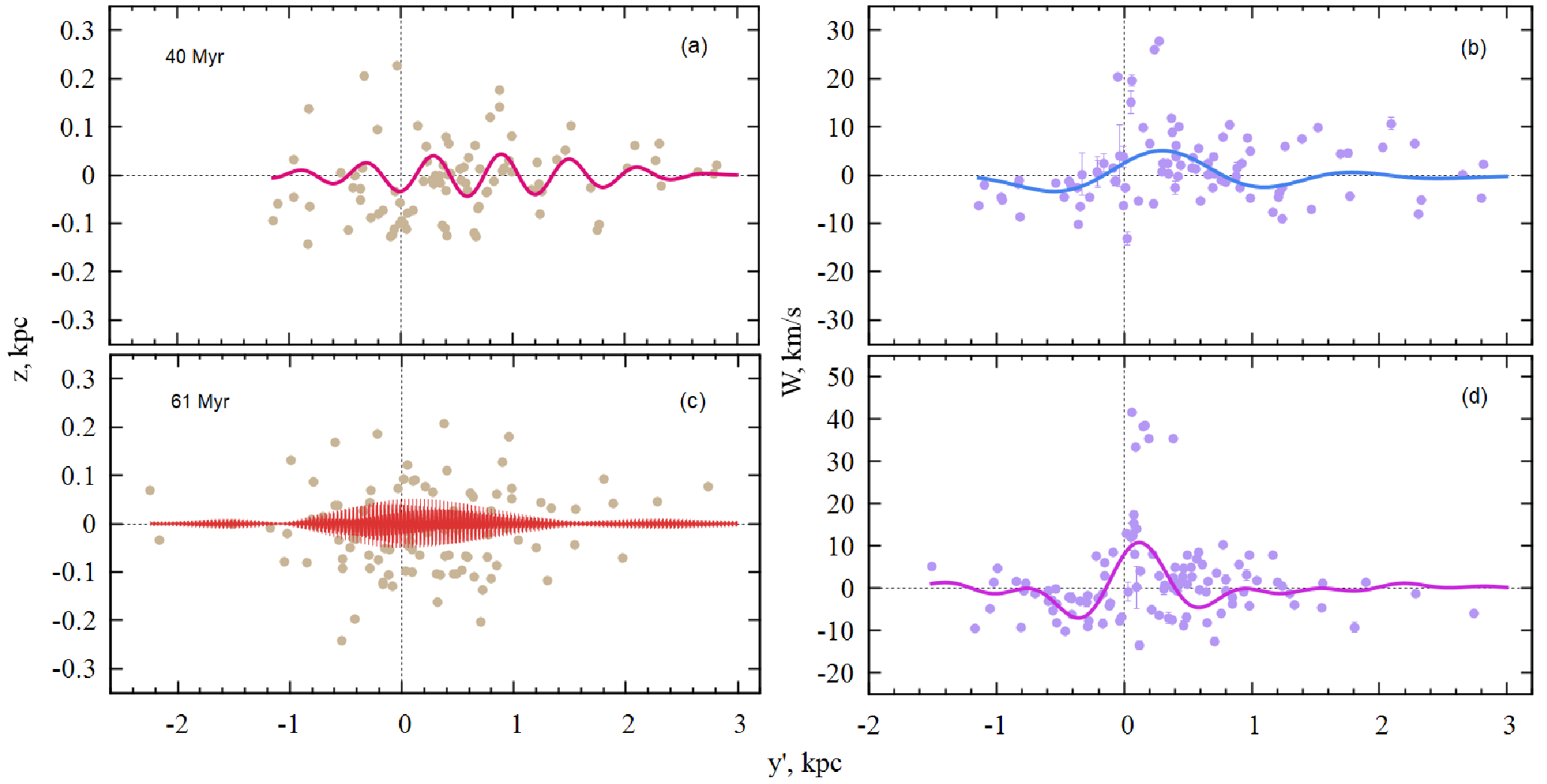}
  \caption{
$z$ coordinates of the OSCs and their vertical velocities $W$ depending on the distance $y'$ from the sample with an average age
of 40~Myr, panels (a) and (b), respectively, and for the OSCs from the sample with an average age of 61~Myr---panels (c) and (d).
 }
 \label{f4-40-61Myr}
\end{center}}
\end{figure}

Note that in our method, the wavelength is determined by the maximum value of the amplitude
spectrum of the analyzed sequence. Figure 2 shows that the maximum value of the spectrum(panel (b)) of
the sequence (panel (a)) corresponds to a wavelength greater than 4 kpc (more precisely, 4.55 kpc which is
indicated by the vertical line in panel (b)). In Fig. 2a, we have added a monochromatic wave (in gray) with
this wavelength. The wave corresponding to the spectrum lobe with the maximum value and more
accurately describing the data is shown in red. From a comparison of these two curves (gray and red), it
can be seen that the wavelengths coincide, since the points of intersection of both curves with the $y'$~axis,
from which the waves begin to move downwards (that is, in the same phase), practically coincide.

The Fourier transform of the vertical velocities shows, in addition to the peak at a wavelength greater than 4 kpc, a much more significant peak at a wavelength of 1.74 kpc, which determines the finer structure of the vertical velocity oscillations. In Fig. 2c, d,
the same lines are plotted as in Fig. 2a, b. From the given spectrum it is evident that the part of the spectrum synchronous with the previous case is not dominant, therefore, the wavelength of the velocity sequence is determined by the peak at a wavelength
of 1.74 kpc.

Based on the coordinates of the OSC sample with an average age of 18.6 Myr, an estimate of the amplitude of the vertical disturbances $z_{max}$  with the wavelength of these disturbances was obtained $\lambda$:
  \begin{equation}
 \label{sol-z18}
 \begin{array}{lll}
   z_{max}= 54\pm5~\hbox{pc},\\
   \lambda= 6.30\pm0.12~\hbox{kpc}.
 \end{array}
 \end{equation}
and according to their velocities---the following estimates:
  \begin{equation}
 \label{sol-W18}
 \begin{array}{lll}
   W_{max}= 7.90\pm0.16~\hbox{km s$^{-1}$},\\
   \lambda= 0.83\pm0.11~\hbox{kpc},
 \end{array}
 \end{equation}
The results of the spectral analysis of the positions and velocities of these OSCs are shown in Fig. 3.

The vertical coordinates of two samples of OSCs with average ages of 40~Myr and 61~Myr, as well as
their vertical velocities $W$ depending on the distance $y'$ are given in Fig. 4. For these samples, we do not
provide spectra in order to save space.

Based on the sample of OSCs with an average age of 40 Myr, the following estimates were obtained: $z_{max} = 43 \pm 2$~pc with the wavelength $\lambda=0.61\pm0.02$~kpc, the amplitude of the vertical velocity disturbance $W_{max}= 5.07\pm0.14$~km s$^{-1}$, and the wavelength found from the vertical velocities is $\lambda=1.78\pm0.04$~kpc.

Based on the OSC sample with an average age of 61 Myr, the following was found: $z_{max} = 51\pm2$~pc
with the wavelength $\lambda=0.02\pm0.02$~kpc, $W_{max}=10\pm2$~km s$^{-1}$, and the wavelength found
from the vertical velocities is $\lambda=1.05\pm0.03$~kpc.

   \section{DISCUSSION}
   \subsection{Spectral Analysis Results}
 Themain criterion for associating the OSCs under study with the Radcliffe wave is the nature of the wave
in their vertical coordinates. There are no questions about the sample of the youngest OSCs (Fig. 2)---the
dependence of their $z$ coordinates on $y'$ has practically the same shape as that of the molecular clouds in the
work of Alves et al. (2020). Even the dependence of the vertical velocities $W$ on $y'$ of this OSC sample
(Fig. 2c) is similar, although, not completely identical.

Parameters (7) are in good agreement with those found for younger OSCs (5). Therefore, we believe
that OSCs with ages in the range of 10--30 Myr are associated with the Radcliffe wave. The distribution
of vertical velocities in Fig. 3c differs from the data in Fig. 2c. In Fig. 3c, we see a strong surge in positive
vertical velocities in the immediate vicinity of the Sun (within a radius of smaller than 200 pc). This difference is apparently due to other reasons. Moreover, the presence of such a surge is also observed in the vertical velocities of older OSCs (Fig. 4b, d). The
complex velocity spectrum of the sample OSCs with ages ranging from 10 to 30 Myr is due to various factors. It has three maxima (Fig. 3d), one of which gives a wavelength of about 2 kpc, the other 4 kpc, which brings this spectrum closer to the spectrum of a
younger sample (Fig. 2d). Therefore, we believe that the large number of OSCs in this sample is closely related to the Radcliffe wave.

The predominance of high-frequency harmonics in the dependences of $z$ on $y'$ in Fig. 4a, c makes it obvious that the OSC samples with average ages of 40 Myr and 61 Myr are not associated with the Radcliffe wave.

Ultimately, we can conclude that the Radcliffe wave is associated with OSCs no older than 30 Myr.

        \subsection{Hypotheses for the RadcliffeWave Origin}
  \subsubsection{Supernova Explosions and the RadcliffeWave}
A recent paper by Marchal and Martin (2023) hypothesized that the Radcliffe wave is the result of
shock fronts generated at the outer boundary of the North Polar Spur affecting gas and dust clouds.

It is known that the Sun is located almost in the center of the Local Bubble, the outer boundary of
which extends no further than 100 pc from the Sun. The formation of the Local Bubble is closely related to
the evolution of stars in the Sco--Cen OB association (Breitschwerdt and de Avillez, 2006; Schulreich et
al., 2018). For example, Fuchs et al. (2006) showed that the Local Bubble began to form 10--15 Myr ago, and since then 14--20 massive stars have exploded as supernovae in the Sco--Cen association.

O'Neill et al. (2024) constructed a three-dimensional model of the magnetic field of the Local Arm. The
authors showed that in the local solar neighborhood,
the magnetic field of the Galaxy is oriented parallel
to the Radcliffe wave. It was also found that the
Local Bubble is more elongated in this direction, and
the molecular clouds on its shell are predominantly concentrated in the perpendicular direction.

Compared to the Local Bubble, the North Polar Spur is a greater phenomenon. At the same time,
there are significantly different points of view regarding the location of the North Polar Spur. Some
authors suggest that the geometric center of this
bubble lies at a distance of 150--200 pc from the Sun (Lallement, 2023). Other authors place the center of the North Polar Spur significantly further away---at a distance of 3--kpc from the Sun (Churazov et al.,
2024) or even at the center of the Galaxy (Sofue and Kataoka, 2021).

Using the coordinates of the two youngest samples of OSCs, we determined the average lines,
along which these clusters are concentrated. The line parameters were determined by the least squares
method. Figure 5 shows the results. From this figure,
we can see that the samples of OSCs of different
ages are shifted relative to each other by a small
distance. In our opinion, such an arrangement does
not contradict the hypothesis about the connection of the Radcliffe wave with the shock wave front that
arose after supernova explosions in the Sco--Cen OB association.

\begin{figure}[t]
{ \begin{center}
  \includegraphics[width=0.65\textwidth]{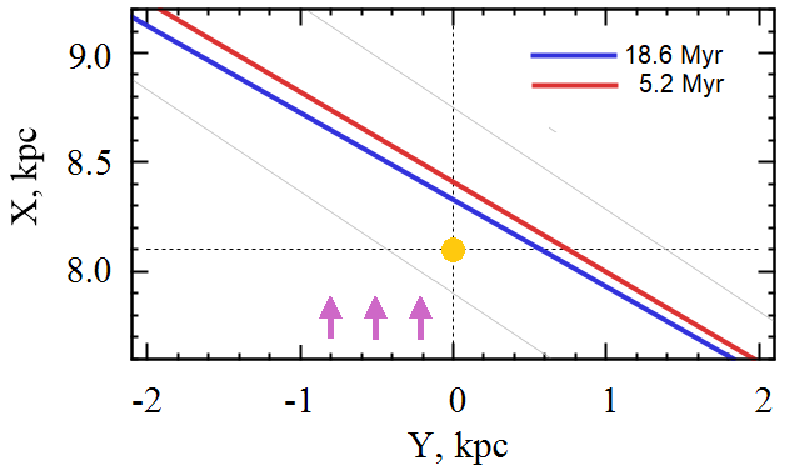}
  \caption{
Average position of two samples of OSCs, the yellow circle marks the position of the Sun, the arrows indicate the
direction of movement of the shock wave fronts formed after supernova explosions in the Sco--Cen OB association.
 }
 \label{f5-model}
\end{center}}
\end{figure}
\begin{figure}[t]
{ \begin{center}
  \includegraphics[width=0.85\textwidth]{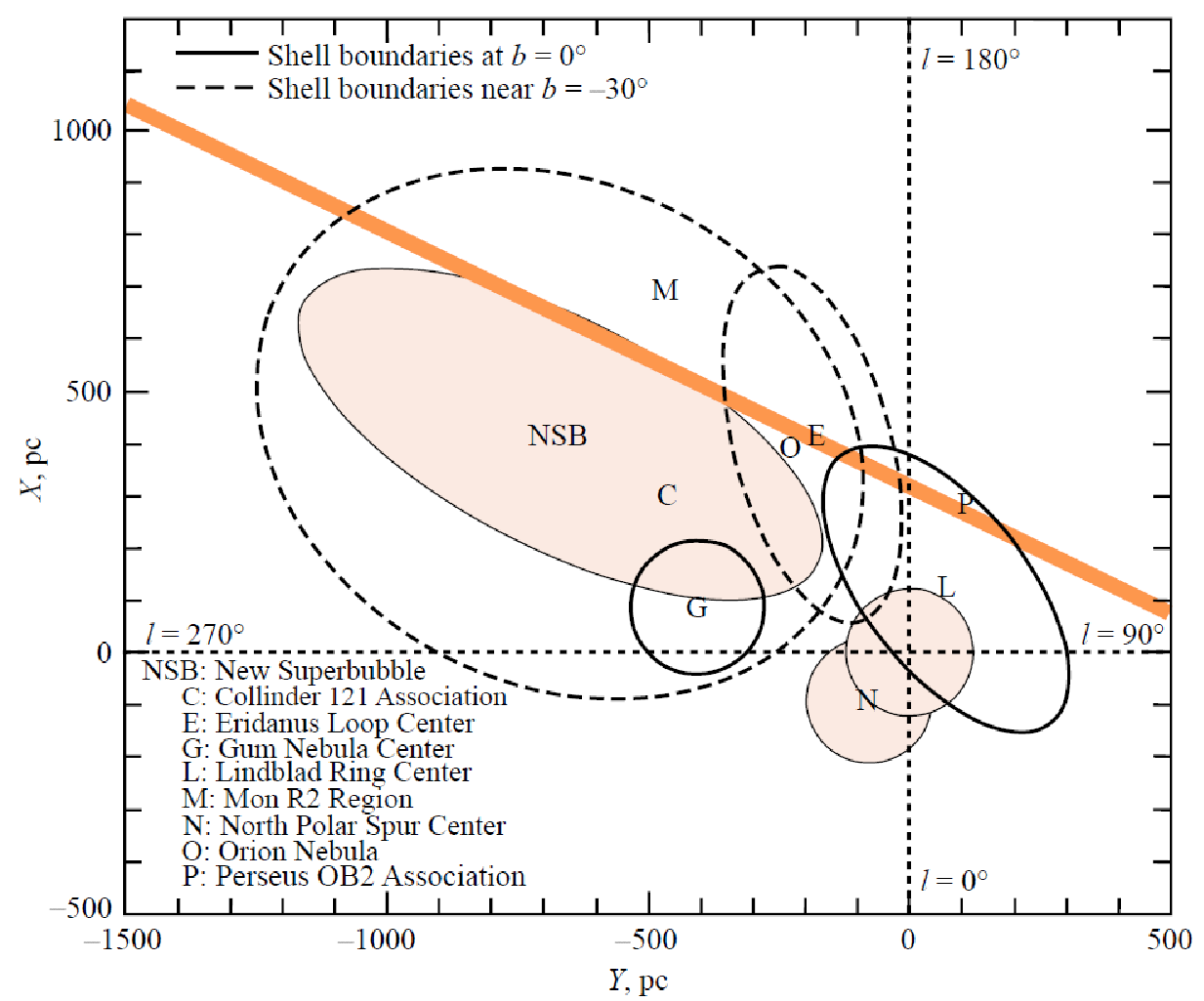}
  \caption{
Position of bubbles, supernova remnants, and a number of young OB associations in the wide solar region. The orange line
shows the position of the Radcliffe wave found in this paper for a sample of the youngest OSCs. The main bubbles are highlighted
in shaded areas, namely, the Local one with the Sun at the center, the North Polar Spur (N), and the New SuperBubble (NSB) first
described by Heiles (1998). We adapted the figure according to Fig. 8 from Heiles (1998).
 }
 \label{f6-Heiles}
\end{center}}
\end{figure}

The arrows in Fig. 5 roughly show the direction of movement of the supposed shock wave fronts formed
after supernova explosions in the Sco--Cen OB association. And the OSCs have located closer to the
epicenters of the explosions were formed first, that is, they have an average age of 18.6 Myr. Then, as the
shock front moved further from the epicenter of the explosions, OSCs with an average age of 5.2 Myr were formed. Here we have an analogy to the age gradient (across the spiral arm) of stars in a spiral density wave.

Thus, there is fundamental agreement with the result from the paper by Konietzka et al. (2024),
where the radial drift of the Radcliffe wave in the Galaxy plane was first detected. Moreover, according
to the data in Fig. 5, we estimated that the velocity of the radial motion of the wave is about 10 pc Myr $^{-1}$
in the region of the Sun (we see that the lines in the figure are not quite parallel).

Figure 6 shows the positions of the centers and outer boundaries of bubbles (regions of low-density
hot gas in the interstellar medium) in the wide circumsolar region according to Heiles (1998). As can
be seen from this figure, there is a close relationship in the location of the centers of the Local Bubble, the
North Polar Spur, theNSB superbubble, and the Eridanus Bubble (marked as ``E'' in the figure) with the
position of the Radcliffe wave. An analysis of Fig. 6 allows us to conclude that the Radcliffe wave may be
the result of the impact on gas and dust clouds of shock waves that arose not only at the outer boundary
of the North Polar Spur, but also at a significantly more extended front comparable in scale to the entire
wave, that is, over a distance of approximately 2 kpc.

  \subsubsection{Magnetic Field and RadcliffeWave}
The hypothesis about the magnetic nature of the Radcliffe wave, based on the scenario of a local emergence of the large-scale magnetic field of the galactic disk due to the so-called Parker instability (Parker, 1966), has recently become the subject of discussion. The results of a study of the magnetic field geometry using polarimetric observations of objects located
in the direction of the Radcliffe wave, presented in Panopoulou et al. (2025), motivated this. The authors
of this paper found that the position angle of linear polarization of radiation from objects located in the
direction of the Radcliffe wave depends significantly on the distance to these sources. In particular, the polarization angle of stars located in the 600-pc vicinity of the Radcliffe wave repeats its spatial structure with
a characteristic inclination to the Galaxy plane at an
angle, the average value of which reaches $10^\circ-18^\circ$.
In the region of the intersection of the Radcliffe wave
with the Galaxy plane, the inclination angle of the
wave, as well as the position angle of polarization, in a region with a linear size of about 350 pc reaches a
value of $10^\circ-30^\circ$. Moreover, the positional angles of radiation polarization of stars located at a distance of
more than 2 kpc are constant and mainly lie in the galactic plane.

The coincidence of the direction of the polarization angle and the direction of the ``ridge'' of the Radcliffe
wave, according to Panopoulou et al. (2025), is a serious argument for studying the hypothesis about the
magnetic nature of this phenomenon. They note that
the parameters of the Radcliffe wave correspond to the
expected spatial and temporal scales of disturbances caused by the Parker instability of the magnetic field in the galactic disk, provided that there is an energetic trigger in the form of a supernova explosion. It should also be noted here that the kinetic energy of the substance forming the Radcliffe wave,
\begin{equation}
 K_{\rm rw} \simeq 10^{51}\,\hbox{erg} \times
  \left(\frac{M_{\rm rw}}{10^6 {\rm M_{\odot}}}\right)
  \left(\frac{v_{\rm rw}}{10^6\,\hbox{cm\,s$^{-1}$}}\right)^2,
 \end{equation}
corresponds to a magnetic field energy of about $\sim 10^{-6}$~G in a galactic disk of about the 300-pc
thickness on a scale of about 2 kpc. This allows us to consider the problem of local disturbance of
the matter of the galactic disk within the framework of the scenario of development of its magnetic field instability without involving additional assumptions about the source of energy of these disturbances (see Kaplan and Pikel'ner, 1979 and references therein).

   \section{Conclusions}
Based on the open star cluster catalog (Hunt and Reffert, 2023), four samples were formed with ages in the intervals of 0--10, 10--30, 30--50, and 50--70~Myr. A selection of potentially Radcliffe wave-related OSCs was carried out from a narrow selection
region inclined to the galactic axis $y$ at an angle of 25$^\circ$. The spectral analysis of the vertical positions and velocities of the selected OSCs showed that the Radcliffe wave is associated with OSCs no older than 30~Myr.

The following estimates of the Radcliffe wave characteristics were obtained for the OSCs with an average age of 5.2~Myr: $z_{max}=117\pm12$~pc with the wavelength $\lambda=4.55\pm0.14$~kpc, and the vertical velocity disturbance amplitude $W_{max}=4.86\pm0.19$~km s$^{-1}$ and the wavelength $\lambda=1.74\pm0.08$~kpc found from vertical velocities.

Based on the sample ofOSCs with an average age of 18.6 Myr (in the range of 10--30 Myr), the following estimates were obtained: $z_{max}= 54\pm5$~pc with the wavelength $\lambda= 6.30\pm0.12$~kpc, the vertical velocity disturbance amplitude
$W_{max}= 7.90\pm0.16$~km s$^{-1}$ with the wavelength $\lambda= 0.83\pm0.11$~kpc. Moreover,
in the vertical velocities of OSCs older than 10 Myr, a strong surge of positive velocities is observed in
the immediate vicinity of the Sun (within a radius of smaller than 200 pc) which greatly influences the result of the spectral analysis of vertical velocities. This surge appears to be due to reasons unrelated to the emergence of the Radcliffe wave.

The radial motion (away from the galactic center) of the Radcliffe wave has been confirmed. The velocity of such movement is 10 pc (Myr)$^{-1}$. In our opinion, the spatial distribution of OSCs of different ages (not older than 30 Myr) does not contradict the hypothesis of the association of the Radcliffe wave with the impact of shock waves from supernova explosions that arose not only on the outer boundary of the North Polar Spur, but on a significantly more extended front comparable in scale to the entire wave, that is, with an approximate size of 2 kpc.

 \subsubsection*{Acknowledgments}
The authors are grateful to the reviewer for useful comments that contributed to improving the paper.

 \subsubsection*{FUNDING}
This work was supported by the organization's
budget. No additional grants to carry out or direct
this particular research were obtained.

 \subsubsection*{CONFLICT OF INTEREST}
The authors of this work declare that they have no
conflicts of interest.

 \subsubsection*{REFERENCES} \small

\quad~1. J. Alonso-Santiago, A. Frasca, A. Bragaglia, et al., Astron. and Astrophys. 691, id. A317 (2024).
DOI:10.1051/0004-6361/202452204

2. J. Alves, C. Zucker, A. A. Goodman, et al., Nature 578 (7794), 237 (2020). DOI:10.1038/s41586-019-1874-z

3. V. V. Bobylev and A. T. Bajkova, Research in Astron. and Astrophys. 24 (3), id. 035010 (2024).
DOI:10.1088/1674-4527/ad113f

4. V. V. Bobylev, A. T. Bajkova, and Y. N. Mishurov, Astronomy Letters 48 (8), 434 (2022).
DOI:10.1134/S1063773722070027

5. D. Breitschwerdt and M. A. de Avillez, Astron. and Astrophys. 452 (1), L1 (2006).
DOI:10.1051/0004-6361:20064989

6. T. Cantat-Gaudin, F. Anders, A. Castro-Ginard, et al., Astron. and Astrophys. 640, id. A1 (2020).
DOI:10.1051/0004-6361/202038192

7. E. Churazov, I. I. Khabibullin, A. M. Bykov, et al., Astron. and Astrophys. 691, id. L22 (2024).
DOI:10.1051/0004-6361/202451762

8. J. Donada and F. Figueras, arXiv eprints aastro/ph:2111.04685 (2021).
DOI:10.48550/arXiv.2111.04685

9. G. Edenhofer, C. Zucker, P. Frank, et al., Astron. and Astrophys. 685, id. A82 (2024).
DOI:10.1051/0004-6361/202347628

10. R. Fleck, Nature 583 (7816), E24 (2020).
DOI:10.1038/s41586-020-2476-5

11. B. Fuchs, D. Breitschwerdt, M. A. de Avillez, et al., MNRAS 373 (3), 993 (2006).
DOI:10.1111/j.1365-2966.2006.11044.x

12. C. Heiles, Astrophys. J. 498 (2), 689 (1998).
DOI:10.1086/305574

13. E. L. Hunt and S. Reffert, Astron. and Astrophys. 673, id. A114 (2023).
DOI:10.1051/0004-6361/202346285

14. S. A. Kaplan and S. B. Pikel'ner, Physics of the Interstellar Medium (Izdatel'stvo Nauka, Mpscow, 1979).

15. R. Konietzka, A. A. Goodman, C. Zucker, et al., Nature 628 (8006), 62 (2024).
DOI:10.1038/s41586-024-07127-3

16. R. Lallement, Comptes Rendus. Physique 23 (S2), 1 (2023). DOI:10.5802/crphys.97

17. R. Lallement, J. L. Vergely, C. Babusiaux, and N. L. J. Cox, Astron. and Astrophys. 661, id. A147
(2022). DOI:10.1051/0004-6361/202142846

18. G.-X. Li and B.-Q. Chen, MNRAS 517 (1), L102 (2022).
DOI:10.1093/mnrasl/slac050

19. G.-X. Li, J.-X. Zhou, and B. Chen, Research Notes Amer. Astron. Soc. 8 (12), id. 316 (2024).
DOI:10.3847/2515-5172/ada0bf

20. A. Marchal and P. G. Martin, Astrophys. J. 942 (2), id. 70 (2023).
DOI:10.3847/1538-4357/aca4d2

21. T. J. O'Neill, A. A. Goodman, J. D. Soler, et al., arXiv e-prints astro/ph:2410.17341 (2024).
DOI:10.48550/arXiv.2410.17341

22. G. V. Panopoulou, C. Zucker, D. Clemens, et al., Astron. and Astrophys. 694, id. A97 (2025).
DOI:10.1051/0004-6361/202450991

23. E. N. Parker, Astrophys. J. 145, 811 (1966). DOI:10.1086/148828

24. M. M. Schulreich, D. Breitschwerdt, J. Feige, and C. Dettbarn, Galaxies 6 (1), id. 26 (2018).
DOI:10.3390/galaxies6010026

25. Y. Sofue and J. Kataoka, MNRAS 506 (2), 2170 (2021).
DOI:10.1093/mnras/stab1857

26. L. Thulasidharan, E. D'Onghia, E. Poggio, et al., Astron. and Astrophys. 660, id. L12 (2022).
DOI:10.1051/0004-6361/202142899

27. A. Vallenari et al. (Gaia Collab.), Astron. and Astrophys. 674, id. A1 (2023).
DOI:10.1051/0004-6361/202243940

28. Z.-K. Zhu, M. Fang, Z.-J. Lu, et al., Astrophys. J. 971 (2), id. 167 (2024).
DOI:10.3847/1538-4357/ad66cd

29. C. Zucker, J. Alves, A. Goodman, et al., ASP Conf. Ser., 534, 43 (2023).
DOI:10.48550/arXiv.2212.00067

\end{document}